\documentclass[prd,amssymb,eqsecnum,showpacs,preprint]{revtex4}
\usepackage{epsfig}
\newcommand{\beq}{\begin{equation}}
\newcommand{\eeq}{\end{equation}}
\newcommand{\bqa}{\begin{eqnarray}}
\newcommand{\eqa}{\end{eqnarray}}
\newcommand{\fr}{\frac}

\newcommand{\fni}{${\mathcal J}^{+}$}

\begin{document}
\title{Absence of trapped surfaces and singularities in cylindrical collapse}
\thanks{To appear in Phys. Rev. D, April 15 issue (in press)}
\author{S\'{e}rgio M. C. V. Gon\c{c}alves}
\affiliation{Department of Physics, Yale University, New Haven, Connecticut 06511}
\date{\today}
\begin{abstract}
The gravitational collapse of an infinite cylindrical thin shell of generic matter in an otherwise empty spacetime is considered. We show that geometries admitting two hypersurface orthogonal Killing vectors cannot contain trapped surfaces in the vacuum portion of spacetime causally available to geodesic timelike observers. At asymptotic future null infinity, however, congruences of outgoing radial null geodesics become marginally trapped, due to convergence induced by shear caused by the interaction of a transverse wave component with the geodesics. The matter shell itself is shown to be always free of trapped surfaces, for this class of geometries. Finally, two simplified matter models are analytically examined. For one model, the weak energy condition is shown to be a necessary condition for collapse to halt; for the second case, it is a sufficient condition for collapse to be able to halt.
\end{abstract}
\maketitle

\section{Introduction}

The issue of formation of horizons in generic gravitational collapse remains an outstanding problem of classical general relativity. The original concept of {\em trapped surface}, due to Penrose~\cite{penrose68}, defines it as a compact spatial two-surface $S$ on which $\theta_{-}\theta_{+}>0$, where $\theta_{\pm}$ are the expansions in the future-pointing null directions orthogonal to $S$. Trapped surfaces signal thus the boundary of a region where any initially expanding null congruence begins to converge; clearly, they define regions of `no escape'. Assuming that cosmic censorship holds, the existence of trapped surfaces implies that of an event horizon, which contains the latter~\cite{hawking&ellis73}. In spherical symmetry, both sufficient and necessary conditions for the occurrence of trapped surfaces have been obtained, which are essentially of the form $m\gtrsim r$, where $m$ and $r$ refer to local (or quasi-local) definitions of mass and radius~\cite{bizon&malec&omurchadha88}. 

Understandably, much less is known about horizon formation in non-spherical geometries. Arguably one of the strongest results to date is that of Shoen and Yau~\cite{shoen&yau83}, who obtained a sufficient criterion for the formation of trapped surfaces in an {\em arbitrary} spacetime: for a given lower bound for the mass density, there is an upper bound for the matter radius, above which trapped surfaces will form. However, being a sufficient condition for the occurrence of trapped surfaces, this result cannot say anything about the conditions under which the collapsing spacetime fails to develop horizons, possibly leading to naked singularities. As with several other notable issues in relativity, a conjecture has been put forward regarding its solution: Thorne's {\em hoop conjecture} states that ``horizons form when and only when a mass $M$ gets compacted into a region whose circumference in {\em every} direction is ${\mathcal C}\lesssim 4\pi M$''~\cite{thorne72}. As originally stated, the conjecture leaves ample room for different definitions of horizon, mass and circumference. In spite of this ambiguity, no known counter-example appears to exist: numerical simulations of prolate and oblate collapse~\cite{shapiro&teukolsky91}, gravitational radiation emission in aspherical collapse~\cite{nakamura&shibata&nakao93}, and analytical studies of prolate collapsing spheroids~\cite{barrabes&israel&letelier91-pelath&tod&wald98} either confirmed or could not refute the conjecture. Detailed analyses of cylindrical pressureless dust~\cite{thorne65phd,echeverria93}, and counter-rotating dust collapse~\cite{apostolatos&thorne92} also confirmed the non-occurrence of horizons.

Cylindrical spacetimes (defined below) constitute an obvious class of spacetimes to study non-perturbative departures from spherical symmetry. Although such spacetimes do not model exactly the dynamical evolution---and inherent gravitational-wave emission---of bounded bodies, they possess a non-trivial field content and constitute thus a valuable test-bed for numerical relativity~\cite{inverno97}, quantum gravity~\cite{ashtekar&pierri96-korotkin&samtleben98}, and for probing the cosmic censorship and hoop conjectures~\cite{berger&chrusciel&moncrief95}. Since several definitions of cylindrical symmetry exist, one must adopt one to work with. In this paper we shall consider cylindrical spacetimes defined by the existence of two commuting spacelike Killing vector fields (such that the orthogonal spacetime is integrable): one translational ($\partial_{z}$) and the other with closed orbits $(\partial_{\phi})$, where the azimuthal coordinate $\phi$ is to be identified at $0$ and $2\pi$. In addition, we take $\partial_{z}$ and $\partial_{\phi}$ to be hypersurface orthogonal, which implies that the cylindrical waves in the vacuum regions admit only one polarization~\cite{beck25,einstein&rosen37}.

The purpose of this paper is twofold: (i) study the occurrence of trapped surfaces in dynamical cylindrical spacetimes with a thin shell of arbitrary matter, and (ii) motivate the notion that realistic (in a suitably defined sense) matter is {\em necessary} to prevent the formation of curvature singularities. Specifically, we consider an infinite cylindrical thin shell with the most general surface stress-energy tensor defined on it, in an otherwise vacuum spacetime. We examine three alternative criteria for trapped surfaces, and show that they can never form, regardless of the matter content for the shell. We also show that---for one class of matter models---the violation of the weak energy condition (WEC) implies that collapse cannot be halted by an outward pressure gradient, and therefore the formation of an infinite spindle-like singularity is inevitable.

The paper is organized as follows: Section II derives the needed mathematical framework, and the vacuum and junction condition equations for the spacetime. Section III discusses trapped surfaces according to three alternative criteria, and shows that there are no trapped cylinders anywhere on the spacetime, except at the limiting case of asymptotic null infinity. This limiting behavior is then explained via the Newman-Penrose formalism. In Sec. IV, we consider two different classes of matter content, and show that, the WEC is either necessary for collapse to halt, or sufficient for it to have a chance of being halted. Section V concludes with a summary and discussion.

Natural geometrized units, in which $8\pi G=c=1$, are used throughout.

\section{Infinite cylindrical thin shell of generic matter}

The complete four-dimensional spacetime consists of an interior vacuum region ${\mathcal M}_{-}$ connected to an exterior vacuum ${\mathcal M}_{+}$ by a three-dimensional thin shell $\Sigma$. The vacuum regions are characterized by the Einstein-Rosen metric~\cite{einstein&rosen37}
\beq
ds^{2}_{\pm}=e^{2(\gamma_{\pm}-\psi_{\pm})}(-dt^{2}_{\pm}+dr^{2}_{\pm})+e^{2\psi_{\pm}}dz^{2}+r^{2}_{\pm}e^{-2\psi_{\pm}}d\phi^{2}, \label{erm}
\eeq
where $\gamma=\gamma(t,r)$, $\psi=\psi(t,r)$, and the coordinate systems $\{x_{\pm}^{\mu}\}$ are adopted. On $\Sigma$ there is a natural holonomic basis $\{e_{(a)}\}$ given by
\beq
e^{\mu}_{(a)}|_{\pm}=\fr{\partial x_{\pm}^{\mu}}{\partial \xi^{a}},
\eeq
where $\{\xi^{a}, a=0,1,2\}$ are intrinsic coordinates on $\Sigma$.
The induced three-metric $\gamma_{ab}$ on $\Sigma$ is then
\beq
\gamma_{ab}=g_{\mu\nu}e^{\mu}_{(a)}e^{\nu}_{(b)},
\eeq
and it is the same on both sides of $\Sigma$, since the four-metric must be continuous across it. This leads to
\bqa
ds^{2}_{\Sigma}&=&-d\tau^{2}+e^{2\psi_{\Sigma}(\tau)}dz^{2}+R^{2}(\tau)e^{-2\psi_{\Sigma}(\tau)}d\phi^{2}, \label{gshell} \\
%\psi^{\Sigma}_{+}(t_{+})&=&\psi_{-}^{\Sigma}(t_{-})=\psi_{\Sigma}(\tau), \\
%r^{\Sigma}_{+}&=&r^{\Sigma}_{-}=R(\tau), \\
\fr{dt_{\pm}}{d\tau}&=&\sqrt{\dot{R}^{2}+e^{2(\psi_{\Sigma}-\gamma_{\pm})}}\equiv\eta_{\pm}. \label{eta}
\eqa
The $zz$ and $\phi\phi$ components of the metric junction condition $[g_{\mu\nu}]\equiv g_{\mu\nu}^{+}-g_{\mu\nu}^{-}=0$ imply that the $r$ coordinate is continuous across $\Sigma$. In the equations above, $\tau$ is the proper time measured by an observer comoving with the shell, with four-velocity 
\beq
u^{\mu}_{\pm}=\eta_{\pm}\delta^{\mu}_{t_{\pm}}+\dot{R}\delta^{\mu}_{r}, \label{fv}
\eeq
where $\dot{R}=dR/d\tau$. The shell $\Sigma$ is defined by
\beq
\Phi(x^{\mu})=r-r_{\Sigma}(t_{\pm})=r-R(\tau)=0.
\eeq
The spacelike unit normal to $\Sigma$ is
\beq
n^{\mu}_{\pm}=\alpha^{-1}(\xi^{a})g^{\mu\nu}_{\pm}\partial^{\pm}_{\nu}\Phi=\dot{R}\delta^{\mu}_{t_{\pm}}+\eta_{\pm}\delta^{\mu}_{r}, \label{sun}
\eeq
where $\alpha\equiv e^{2(\psi_{\Sigma}-\gamma_{\pm})}\eta^{-1}_{\pm}$ is a normalization factor.

The normal extrinsic curvature, $K_{ab}^{\pm}$, is~\cite{israel66-67}
\beq
K_{ab}:=-n_{\mu}e^{\nu}_{(b)}\nabla_{\nu}e^{\mu}_{(a)}=-n_{\sigma}\left(\fr{\partial^{2}x^{\sigma}}{\partial\xi^{a}\partial\xi^{b}}+\Gamma^{\sigma}_{\mu\nu}\fr{\partial x^{\mu}}{\partial\xi^{a}}\fr{\partial x^{\nu}}{\partial\xi^{b}}\right).
\eeq
Its non-vanishing components are:
\bqa
K^{\pm}_{\tau\tau}&=& D_{\perp}\psi_{\pm}-D_{\perp}\gamma_{\pm}-\fr{1}{\eta_{\pm}}\left[\ddot{R}+\dot{R}(\dot{\gamma}_{\pm}-\dot{\psi}_{\Sigma})\right], \\
K_{zz}^{\pm}&=& e^{2\psi_{\Sigma}}D_{\perp}\psi_{\pm}, \\
K_{\phi\phi}^{\pm}&=& e^{-2\psi_{\Sigma}}R(-\eta_{\pm}+RD_{\perp}\psi_{\pm}),
\eqa
where $D_{\perp}\equiv n^{\mu}\nabla_{\mu}$ is the normal derivative with respect to $\Sigma$, and the object
\beq
\fr{d^{2}r_{\Sigma}(t_{\pm})}{dt^{2}_{\pm}}=\eta^{-4}_{\pm}e^{2(\psi_{\pm}-\gamma_{\pm})}[\ddot{R}+\dot{R}(\dot{\gamma}_{\pm}-\dot{\psi}_{\Sigma})]
\eeq
was used.

For the matter content on the shell, we consider the most general stress-energy tensor $S_{ab}$ defined on $\Sigma$, compatible with the metric (\ref{gshell}):
\beq
S_{ab}=\rho\delta^{\tau}_{a}\delta^{\tau}_{b}+p_{z}e^{2\psi_{\Sigma}}\delta^{z}_{a}\delta^{z}_{b}+p_{\phi}R^{2}e^{-2\psi_{\Sigma}}\delta^{\phi}_{a}\delta^{\phi}_{b},
\eeq
where $\rho$, $p_{z}$, and $p_{\phi}$ are the proper surface energy density, pressure along the $z$-direction, and surface stress, respectively, as measured by an observer comoving with the shell with four-velocity given by (\ref{fv}).

\subsection{Vacuum field equations}

The vacuum field equations are (where the $\pm$ subscript has been dropped for simplicity):
\bqa
&&\psi_{,tt}-r^{-1}\psi_{,r}-\psi_{,rr}=0, \label{weq} \\
&&\gamma_{,t}=2r\psi_{,r}\psi_{,t}, \label{e1} \\
&&\gamma_{,r}=r[(\psi_{,t})^{2}+(\psi_{,r})^{2}]. \label{e2}
\eqa
We are thus left with the axisymmetric wave equation (\ref{weq}), which is {\em decoupled} from Eqs. (\ref{e1})-(\ref{e2}). One first solves the wave equation with appropriate boundary data, and then solve for (\ref{e1})-(\ref{e2}) by quadratures. Remarkably, these two equations are compatible because their integrability condition is precisely the wave equation (\ref{weq}). 

The general (outgoing) solution to the wave equation (\ref{weq}) is given by
\beq
\psi(t,r)=\Re \int^{\infty}_{0} A(\omega)e^{-i\omega t}H_{0}^{(1)}(\omega r)d\omega,
\eeq
where $A(\omega)$ is a complex-valued function, and $H_{n}^{(1)}(x):=J_{n}(x)+iY_{n}(x)$ is a Hankel function of the first kind, and $J_{n}$ and $Y_{n}$ are Bessel functions of the first and second kind, respectively~\cite{abramovitz&stegun64}.

\subsection{Junction conditions for the shell}

On the shell, Einstein's equations reduce to the Darmois-Israel junction conditions for the normal extrinsic curvature~\cite{darmois27,israel66-67}:
\beq
[K_{ab}]\equiv K_{ab}^{+}-K_{ab}^{-}=-S_{ab}+\fr{1}{2}S\gamma_{ab}.
\eeq
The non-vanishing components of the above equation are $zz$, $\phi\phi$, and $\tau\tau$, which yield, respectively:
\bqa
[D_{\perp}\psi] &=&-\fr{1}{2}\Delta, \label{jcz} \\
\left[\eta\right] &=& -R\rho, \label{jcph} \\
\left[\eta\gamma_{,r}\right]&=&p_{\phi}-\fr{R\rho}{\eta_{+}\eta_{-}}(\ddot{R}-\dot{R}\dot{\psi}_{\Sigma}), \label{ktt}
\eqa
where 
\beq
\Delta\equiv2p_{z}-S^{a}_{a}=\rho+p_{z}-p_{\phi}.
\eeq
Using the field equations (\ref{e1})-(\ref{e2}), Eq. (\ref{ktt}) can be cast solely in terms of the wave-field $\psi$:
\beq
\ddot{R}=\dot{R}\dot{\psi}_{\Sigma}+\eta_{+}\eta_{-}\fr{p_{\phi}}{R\rho}-R\{\dot{\psi}^{2}_{\Sigma}+(D_{\perp}\psi_{-})^{2}\}-\fr{\eta_{-}}{\rho}\Delta\left(\fr{\Delta}{4R}-D_{\perp}\psi_{-}\right). \label{jctau}
\eeq
Equations (\ref{jcz}) and (\ref{jcph}) match the metric functions $\psi$ and $\gamma$ across $\Sigma$, and Eq. (\ref{jctau}) governs the motion of the shell. We note that, in the absence of a matter shell ($\rho=0$), it immediately follows from Eqs. (\ref{eta}) and (\ref{jcph}) that the `time' coordinate $t$ is continuous across $\Sigma$ and hence it can be globally defined, as expected.

An additional constraint is obtained by the conservation of energy-momentum on $\Sigma$:
\beq
^{(3)}\nabla_{b}S^{b}_{a}=-[e^{\mu}_{(a)}T_{\mu\nu}n^{\nu}],
\eeq
the only non-vanishing component of which gives
\beq
\dot{\rho}+(\rho+p_{\phi})\fr{\dot{R}}{R}+(p_{z}-p_{\phi})\dot{\psi}_{\Sigma}=[e^{\mu}_{(\tau)}T_{\mu\nu}n^{\nu}]=0. \label{econ}
\eeq

\section{Trapped surfaces}

\subsection{Proper circumference radius criterion}

We shall use the following operational definition of trapped surface, which is a natural extension of Penrose's~\cite{penrose68}, for surfaces that are not necessarily compact nor two-dimensional: let $S$ be any non-null surface and $\theta_{\pm}$ the expansion in the future-oriented null directions normal to $S$, then $S$ is said to be a trapped surface iff $\theta_{+}\theta_{-}|_{S}\geq0$, where the inequality saturates for the marginally trapped case. In the present case, our surface $S$ is that of an infinite cylinder with proper circumference radius ${\mathcal R}=re^{-\psi}$. 

Modulo the two-dimensional quotient space orthogonal to the symmetry axis, the Einstein-Rosen metric is conformal to Minkowski spacetime (which is why the four-dimensional wave equation is formally the same as that of flat space), and thus it suffices to introduce flat spacetime null coordinates (where, for simplicity, the `$\pm$' subscript is omitted):
\beq
u=t-r,\;\; v=t+r,
\eeq
to study the expansion of $S$ along normal null directions. We have then
\beq
\theta_{\pm}=(\partial_{t}\mp\partial_{r}){\mathcal R}.
\eeq
Trapped surfaces form when
\beq
\theta_{+}\theta_{-}=({\mathcal R}_{,t})^{2}-({\mathcal R}_{,r})^{2}\geq0. \label{tsc}
\eeq

\subsubsection{Vacuum regions}

In the vacuum regions, the above condition reads
\beq
r^{2}\psi_{,t}^{2}\geq(1-r\psi_{,r})^{2}.
\eeq
One can use the asymptotic properties of Bessel functions (cf. Appendix B) to show that, for $\omega r\gg1$,
\beq
\psi_{,t}\sim -\psi_{,r}\sim\fr{1}{\sqrt{r}}\Re \int_{0}^{\infty} \omega A(\omega)e^{-i[\omega(t+r)-3\pi/4]}d\omega. \label{asb}
\eeq
Hence, for $\omega r\gg1$,
\beq
\theta_{+}\theta_{-}\sim-1+2r\psi_{,t}\psi_{,r}<0.
\eeq
For large radii, this regime covers `almost' all the frequencies, whereas for small radii, the validity of the inequality above is restricted to high-frequency modes with $\omega\gg r^{-1}\gg1$.

In general, it proves more convenient to look for {\em outer marginally trapped surfaces} (OMTS)---surfaces whose expansion vanishes along future-oriented null normal directions---since their existence is implied by that of trapped surfaces~\cite{wald84}. Hence, it suffices to show that no OMTS form, to prove that there are no trapped surfaces. The condition for OMTS is
\beq
\theta_{+}=-\psi_{,t}-\fr{1}{r}+\psi_{,r}=0.
\eeq
Partial differentiation of this equation with respect to $t$ and $r$ yields, respectively,
\beq
\psi_{,tr}=\psi_{,tt}, \;\;\;\; \psi_{,tr}=\psi_{,rr}+\fr{1}{r^{2}}.
\eeq
Combining these two equations with the wave equation (\ref{weq}) leads to
\beq
\psi_{,r}=\fr{1}{2}\psi_{,t}=-\fr{1}{r}. \label{mtsc}
\eeq
Now, from Eq. (\ref{asb}), we know that these two equalities fail to hold for $\omega r\gg1$. To show that the same happens for $\omega r\lesssim1$, we note that
\beq
\int_{0}^{a} x^{n}H_{0}^{(1)}(x)dx<\infty \;\;\; \mbox{iff}\;\;\; n>-1, \nonumber
\eeq
from which it follows that, as $\omega\rightarrow0$, $A(\omega)\sim\omega^{n}$ with $n>-1$. Hence, for $\omega r\lesssim1$, we have
\beq
\psi_{,t}\sim\psi_{,r}\sim\fr{1}{r^{2+n}},
\eeq
thereby showing that Eq. (\ref{mtsc}) also fails to hold for $\omega r\lesssim1$. By construction, all of the above holds inside and outside the shell, and we conclude therefore that there are no trapped surfaces in neither vacuum region.

\subsubsection{Shell}

On the shell, the condition for OMTS is
\beq
\theta_{+}=\fr{R}{\eta}e^{-\psi_{\Sigma}}\left(\fr{\dot{R}}{R}-\dot{\psi}_{\Sigma}-\fr{\eta}{R}\right)=0.
\eeq
Since $\eta/R>0$, an obvious {\em necessary} condition for OMTS is
\beq
\fr{\dot{R}}{R}-\dot{\psi}_{\Sigma}>0.
\eeq
However, for a collapsing configuration the physical circumference radius must decrease with proper time: 
\beq
\dot{\mathcal R}<0\Rightarrow \dot{R}-R\dot{\psi}_{\Sigma}<0,
\eeq
which contradicts the necessary condition for OMTS. Hence, a collapsing shell can never become trapped, irrespective of the details of the matter content.

\subsection{Specific area radius criterion}

An alternative definition of trapped cylinder has been recently proposed by Hayward~\cite{hayward00}, and relies on the specific area radius, which is constructed as follows. The two spacelike Killing vectors $\{\partial_{z},\partial_{\phi}\}$ have a well-defined invariant geometric meaning:
\beq
{\mathcal R}:=|\partial_{\phi}|=\sqrt{\partial_{\phi}\cdot\partial_{\phi}}
\eeq
is the proper circumference radius, and
\beq
l:=|\partial_{z}|=\sqrt{\partial_{z}\cdot\partial_{z}}
\eeq
is the specific Killing length. One can thus define a specific area for the cylinder as
\beq
A:=2\pi{\mathcal R}l=2\pi\tilde{r},
\eeq
where $\tilde{r}={\mathcal R}l$ is the specific area radius. A cylinder is said to be trapped, marginal, or untrapped if the vector $N_{\mu}:=\nabla_{\mu}\tilde{r}$ is timelike, null, or spacelike, respectively~\cite{hayward00}.  In our case, we have
\beq
{\mathcal R}=re^{-\psi},\;\; l=e^{\psi},\;\; \tilde{r}=r.
\eeq
From a geometrical viewpoint, this definition is perfectly acceptable, since, modulo invariant rotations, all of the spatial (`radial') submanifold orthogonal to the axis is covered by $r\in[0,+\infty)$. However, from a physical standpoint, this definition is somewhat lacking, since the specific area radius is simply a coordinate radius, and as such {\em changes} under the rescaling $z\rightarrow \alpha z$ as $r\rightarrow \alpha^{-1/2}r$, unlike the proper circumference radius, which remains invariant under such rescaling. In addition, an external timelike observer can only measure proper circumferences, but not coordinate radii. Despite these shortcomings, and because of its useful geometrical meaning, we shall examine the formation of trapped surfaces according to it.

\subsubsection{Vacuum regions}

In the vacuum regions,
\beq
N_{\mu}^{\rm vac}=\delta^{r}_{\mu},
\eeq
and thus
\beq
\left(N_{\mu}N^{\mu}\right)_{\rm vac}=e^{2(\psi-\gamma)}\geq0. \label{nvac}
\eeq
Hence there are no trapped surfaces in the vacuum regions provided the quantity $\psi-\gamma$ does not diverge negatively (in which the case a marginally trapped surface would form). Since $g_{rr}=-g_{tt}=e^{2(\gamma-\psi)}$, it follows that for {\em regular} spacetimes the inequality must be strict and thus there are no trapped surfaces. This is true for {\em any} cylindrical regular metric with two commuting hypersurface orthogonal Killing vectors. Accordingly, it holds inside and outside $\Sigma$. In the interior vacuum region, it is straightforward to show that regularity of the axis implies absence of trapped surfaces. Let $\chi\equiv |\partial_{\phi}|^{2}=r^{2}e^{-2\psi}$; then the symmetry axis is regular iff the following local flatness condition is obeyed~\cite{exactsols}
\beq
\lim_{r\rightarrow0} \fr{\chi_{,\mu}\chi_{,}^{\mu}}{4\chi}=1,
\eeq
which requires
\beq
\gamma(t,0)=0,\;\; \left.\psi_{,t}\right|_{r=0}<\infty, \;\;\left.\psi_{,r}\right|_{r=0}<\infty. \label{regc}
\eeq
Hence, $|N|^{2}_{\rm vac}>0$ along the axis.

\subsubsection{Shell}

On the shell,
\beq
N_{\mu}^{\Sigma}=\dot{R}\eta^{-1}\delta_{\mu}^{t}+\delta_{\mu}^{r},
\eeq
which gives
\beq
\left(N_{\mu}N^{\mu}\right)_{\Sigma}=\eta^{-2}e^{2(\psi-\gamma)}(\eta^{2}-\dot{R}^{2})=\eta^{-2}e^{4(\psi-\gamma)}\geq0,
\eeq
where the junction condition (\ref{eta}) was used in the last equality. Since the metric must be regular on $\Sigma$, the inequality can never saturate, and thus $\Sigma$ is free of trapped surfaces.

\subsection{Geodesic null congruences}

A direct, physical way to probe the occurrence of trapped surfaces in the vacuum spacetime(s), is to examine the divergence of a congruence of outgoing radial (i.e., orthogonal to the symmetry axis) null geodesics (ORNG). To examine the behavior along null directions, it is convenient to introduce retarded Bondi coordinates $\{u,r,z,\phi\}$, in which the Einstein-Rosen metric reads
\beq
ds^{2}=-e^{2(\gamma-\psi)}(du^{2}+2dudr)+e^{2\psi}dz^{2}+r^{2}e^{-2\psi}d\phi^{2}.
\eeq
In these coordinates, the vacuum field equations are
\bqa
&&2\psi_{,ur}-\psi_{,r}+r^{-1}(\psi_{,u}-\psi_{,r})=0, \\
&&\gamma_{,u}=2r\psi_{,u}(\psi_{,u}+\psi_{,r}), \label{ne1} \\
&&\gamma_{,r}=r(\psi_{,r})^{2}. \label{ne2}
\eqa

The null vector field tangent to the ORNG is given by
\beq
k^{\mu}=\fr{dx^{\mu}}{d\lambda}=X(u,r)\delta^{\mu}_{r},
\eeq
where $\lambda$ is an affine parameter along the geodesics, and $X>0$ since the geodesics are outgoing (i.e., $r$ is a monotonically increasing function of the affine parameter). The geodesic equation reads then
\beq
k^{\nu}\nabla_{\nu}k^{\mu}=X[X_{,r}+2X(\gamma_{,r}-\psi_{,r})]\delta^{\mu}_{r}=0,
\eeq
which readily integrates to
\beq
X(u,r)=f(u)e^{2(\psi-\gamma)}, \label{orng}
\eeq
where $f(u)>0$ is an arbitrary function, fixed by the choice of $X$ for a given value of the coordinate $r$.
For such a geodesic congruence, the expansion scalar is
\beq
\Theta:=\nabla_{\mu}k^{\mu}=\fr{X(u,r)}{r}=\fr{f}{r}e^{2(\psi-\gamma)}.
\eeq
Clearly, as long as one restricts oneself to finite values of $r$, there are no trapped surfaces. There could be trapped surfaces at asymptotic future null infinity~\cite{ftn1} \fni if, for fixed $u$,  $\lim_{r\rightarrow+\infty} X\lesssim{\mathcal O}(r)$. From a physical viewpoint, the possible occurrence of marginally trapped surfaces at \fni is not worrisome, in that geodesic timelike observers can never reach \fni~\cite{ftn2}. However, from an asymptotic infinity viewpoint, it is of interest to examine the limiting behavior of geodesic focusing in a radiative spacetime. The asymptotic behavior of $\Theta$ at \fni is given by
\bqa
\Theta|_{{\mathcal J}^{+}}&=&\lim_{r\rightarrow+\infty} \left.\fr{X(u,r)}{r}\right|_{u=u_{0}} \nonumber \\
&=&\lim_{r\rightarrow+\infty} \fr{f(u_{0})}{r}e^{2(\psi(u_{0},r)-\gamma(u_{0},r))}. \label{nexp}
\eqa
The behavior of the wave-field $\psi$ [which determines $\gamma$ via the field equations (\ref{ne1})-(\ref{ne2})] at \fni was computed by Ashtekar, Bi$\check{\rm c}$\'{a}k, and Schmidt~\cite{ashtekar&bicak&schmidt97,ashtekar&bicak&schmidt97b}, and yields
\bqa
\psi(u,r)&=&\fr{c_{0}(u)}{\sqrt{r}}+\fr{1}{\sqrt{r}}\sum_{n=1}^{\infty}c_{n}(u)r^{-n}, \label{pinf} \\
\gamma(u,r)&=&\gamma_{0}-2\int^{u}_{-\infty} c'^{2}_{0} du -\sum_{n=1}^{\infty} \fr{b_{n}(u)}{n+1}\fr{1}{r^{1+n}}, \label{ginf}
\eqa
where $c'_{0}=dc_{0}/du$, and the coefficients $c_{n}$ and $b_{n}$ are determined from the initial Cauchy data.
From Eqs. (\ref{nexp})-(\ref{ginf}) it follows that
\beq
\Theta|_{{\mathcal J}^{+}}=0.
\eeq
This {\em limiting behavior} reflects the fact that, near \fni the cylindrical waves behave locally like plane waves, which intersect the congruence of ORNG, thereby causing them to start (marginally) converging (cf. Fig. 1). 

\begin{figure}
\begin{center}
%\leavemode
\epsfysize=14pc
\epsfxsize=24pc
\epsffile{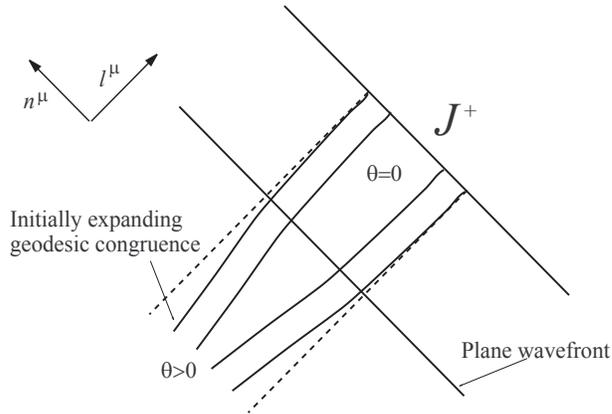}
\end{center}
\caption{Asymptotic behavior of an initially expanding future-oriented null geodesic congruence at \fni. When the radius (proper or otherwise) of cylindrical waves tends to infinity, they become plane waves, whose wavefronts orthogonally intersect the null geodesic congruence. The transverse wave component in the $n^{\mu}$ direction induces shear on the congruence, which in turn makes it focus, by virtue of the Raychaudhuri-type equation (\ref{npexp}). \label{fig1}}
\end{figure}

Such behavior is best described using the Newman-Penrose formalism~\cite{np}. The two relevant equations are (cf. Appendix C)
\bqa
D\theta&=&\theta^{2}+\sigma\bar{\sigma}+\Phi_{00}, \label{npexp} \\
D\sigma&=&\sigma(\theta+\bar{\theta})+\Psi_{(0)}, \label{npshe}
\eqa
which govern the evolution of the expansion $-\theta$ and shear $\sigma$ of the ORNG, respectively. Near \fni the congruence is orthogonally intersected by plane waves aligned with $n^{\mu}$ (cf. Fig. 1). The existence of a transverse wave component in the $n^{\mu}$ direction~\cite{ftn3} is signaled by a non-vanishing $\Psi_{(0)}$:
\beq
\Psi_{(0)}=-C_{\alpha\beta\mu\nu}l^{\alpha}m^{\beta}l^{\mu}m^{\nu}\neq0.
\eeq
This non-zero $\Psi_{(0)}$ implies---by Eq. (\ref{npshe})---that the congruence starts to shear, with the shear axis being determined by the polarization of the wave (i.e., the wave-field $\psi$). This introduces a non-negative term 
\beq
\sigma\bar{\sigma}=\left(\psi_{,t}-\psi_{,r}+\fr{1}{2r}\right)^{2}
\eeq
into Eq. (\ref{npexp}), thereby inducing the congruence to converge.

\subsection{$C$-energy}

The concept of $C$-energy was introduced by Thorne~\cite{thorne65} as a local definition of energy density per unit Killing-length $z$ for cylindrical systems. Being a locally defined quantity in spacetimes with a translational Killing field, its relation to energy-like quantities defined in asymptotically flat spacetimes (e.g., the ADM or Bondi mass) becomes unclear~\cite{ftn4}. In fact, as shown by Ashtekar and Varadarajan~\cite{ashtekar&varadarajan94,varadarajan95}, the Hamiltonian that generates asymptotic time translations at spatial infinity is {\em not} the $C$-energy, but a non-polynomial function of it, which is positive and bounded from above. Nevertheless, being a locally conserved and measurable quantity, $C$-energy remains a very useful tool for the analysis of cylindrical systems, and, unsurprisingly, it can be linked to the occurrence of conical singularities~\cite{xanthopoulos86} and trapped surfaces~\cite{hayward00}.

Following Thorne~\cite{thorne65}, we introduce the $C$-energy potential
\beq
C(t,r):=\fr{1}{8}(1-e^{-2\gamma}), \label{ce}
\eeq
which is proportional to the total $C$-energy contained inside a cylinder of specific area radius $r$ per unit Killing-length $z$. The $C$-energy flux vector $P^{\mu}$ is given by
\bqa
P^{\mu}&:=& \fr{1}{8}\epsilon^{\mu\nu\alpha\beta}\sqrt{-g}[\ln(1-8C)]_{,\nu}\fr{\partial_{(z)\alpha}}{|\partial_{z}|^{2}}\fr{\partial_{(\phi)\beta}}{|\partial_{\phi}|^{2}} \nonumber \\
&=&\fr{e^{2(\psi-\gamma)}}{r}(C_{,r}\delta^{\mu}_{t}-C_{,t}\delta^{\mu}_{r}),
\eqa
and obeys the local conservation law $\nabla_{\mu}P^{\mu}=0$. An observer with 4-velocity $u^{\mu}$ measures a $C$-energy density ${\mathcal E}=P^{\mu}u_{\mu}$, and the $C$-energy flux across a hypersurface $\Sigma$ with spacelike normal $n^{\mu}$ (such that $n^{\mu}u_{\mu}=0$) is ${\mathcal F}=P^{\mu}n_{\mu}$.

From Eqs. (\ref{nvac}) and (\ref{ce}) it follows that
\beq
(N_{\mu}N^{\mu})_{\rm vac}=e^{2\psi}\left(1-8C\right).
\eeq
Hence, trapped surfaces form whenever
\beq
C\geq\fr{1}{8},
\eeq
where the inequality saturates for the marginally trapped case. This is the cylindrical analogue of the well-known asymptotically flat spherically symmetric condition $E\geq1/2$ for Tolman-Bondi spacetimes~\cite{tbeg}, where $E\equiv m_{\rm MS}/R_{\rm p}$ is the Misner-Sharp mass~\cite{misner&sharp64} per unit proper area radius. In the present case, $C<1/8$ and the vacuum regions are untrapped.

\section{Singularities}

The issue of singularity formation in cylindrical collapse of a thin shell has been addressed for the particular cases of pressureless dust and counter-rotating dust (i.e., with a non-vanishing tangential pressure component). The pure dust case was analyzed by Thorne~\cite{thorne65phd} and Echeverria~\cite{echeverria93}, who showed that complete collapse is inevitable, and an infinite spindle-like singularity must thus form as a result. The case of counter-rotating dust coupled to Einstein-Rosen waves was addressed by Apostolatos and Thorne~\cite{apostolatos&thorne92}, who resorted to sequences of momentarily static radiation-free (MSRF) configurations and $C$-energy balance arguments to conclude that an arbitrarily small amount of counter-rotation (and, very likely, also of net rotation) suffices to halt collapse, thereby precluding the formation of spindle singularities. The counter-rotating case was recently revisited by Pereira and Wang~\cite{pereira&wang00}, who considered a flat interior and an outgoing null fluid exterior (as opposed to a `realistic' vacuum containing cylindrical waves); by means of further assumptions for the matter content, they obtained a simple solvable model which admits initial data leading to a spindle singularity, thus in apparent contradiction with the earlier results. However, the model is highly contrived, and the assumptions made render it unphysical, since the dynamics of the exterior null fluid can only be obtained {\em a posteriori}, after a particular solution for the motion of the shell is derived. It is therefore conceivable that a more realistic model, wherein the emission of gravitational radiation arises {\em dynamically} from the inward accelerated motion of the shell, will qualitatively agree with the results of~\cite{apostolatos&thorne92}.

We shall conjecture here~\cite{ftn5} that {\em realistic matter is required to prevent the occurrence of singularities.} To support this conjecture, we examine below two classes of matter models for which the WEC is either necessary for collapse to halt, or sufficient for it to be able to halt.

\subsection{$\Delta=0$}

We consider here the case $\Delta=0\Leftrightarrow\rho=p_{\phi}-p_{z}$. This case includes the Pereira-Wang matter model ($p_{z}=0$, $\rho=p_{\phi}$), and a subcase of the Apostolatos-Thorne model ($p_{z}=0$, but not necessarily $\rho=p_{\phi}$). Then, the junction conditions (\ref{jcz}) and (\ref{jctau}) simplify to
\bqa
[D_{\perp}\psi]&=&0, \\
\ddot{R}&=&\dot{R}\dot{\psi}_{\Sigma}+\eta_{+}\eta_{-}\fr{p_{\phi}}{R\rho}-R\{\dot{\psi}^{2}_{\Sigma}+(D_{\perp}\psi_{-})^{2}\}, \label{dyn}
\eqa
and the local conservation equation on the shell becomes
\beq
\fr{\dot{\rho}}{\rho}+\left(1+\fr{p_{\phi}}{\rho}\right)\fr{\dot{R}}{R}-\dot{\psi}=0.
\eeq
This equation can be easily integrated by assuming a functional relation between $\rho$ and $p_{\phi}$, which, for simplicity, we take to be of barotropic form: $p_{\phi}=\alpha\rho$, where the real constant $\alpha^{2}<1$, to preserve causality. This gives then
\beq
\rho=A\fr{e^{\psi_{\Sigma}}}{R^{1+\alpha}}, \label{ssol}
\eeq
where $A>0$ is an integration constant, which is {\em dimensionful} and scales as $[A]\sim R^{\alpha}\sim L^{\alpha}$, such that $[\rho]\sim L^{-1}$. This constant $A$ turns out to be constrained by the fact that there is an upper limit for the total rest mass per unit proper length, which arises from the requirement of radial nonclosure of space. Specifically, the space around a given cylindrical configuration is radially non-closed iff~\cite{ftn6}
\beq
2\pi R\sigma_{\rm AT}<\fr{1}{4},
\eeq
where $\sigma_{\rm AT}\equiv\rho/(8\pi)$ is the total rest mass per unit proper area (the factor of $8\pi$ arises because we work with units in which $8\pi G=1$, whereas Apostolatos and Thorne use $G=1$). In our case, the total rest mass per unit proper length is the dimensionless quantity
\beq
m_{\rm p}=\fr{1}{4}R\rho=\fr{1}{4}Ae^{\psi_{\Sigma}}R^{-\alpha},
\eeq
from which it follows that we must have 
\beq
R\rho<1 \label{rnc},
\eeq
or $A<e^{-\psi_{\Sigma}}R^{\alpha}$, if (\ref{ssol}) holds. (We note that this constraint on the rest mass per unit proper length plays no role in the subsequent assertion that the WEC is a necessary condition for singularities not to form.)

We now look for local extrema of $R$, and show that the obedience of the WEC is necessary for collapse to be halted. Local extrema of $R$ occur at $\tau=\tau_{\rm e}$, such that $\dot{R}_{\rm e}=0$ and $\ddot{R}_{\rm e}\neq0$. From Eq. (\ref{dyn}) we have then
\beq
\ddot{R}_{\rm e}=-\eta_{-}(\rho+p_{z})\left(1-\fr{\eta_{-}}{R\rho}\right)-R\{\dot{\psi}^{2}_{\Sigma}+(D_{\perp}\psi_{-})^{2}\},
\eeq
where the junction condition (\ref{jcph}) was used, and the right-hand-side is to be evaluated at $\tau=\tau_{\rm e}$. Since the last summand is manifestly negative and $\eta_{-}>0$, a necessary condition for collapse to be halted (i.e., for $\ddot{R}_{\rm e}>0$) is
\beq
(\rho+p_{z})\left(\fr{\eta_{-}}{R\rho}-1\right)>0. \label{ncc}
\eeq
By Eqs. (\ref{eta}) and (\ref{jcph}), the second term is always positive for $\rho>0$. It then follows that we must have $\rho+p_{z}>0$ if collapse is to be halted. This means that the WEC (which requires $\rho>0$ and $\rho+p_{i}>0\,\forall\,i$) must be satisfied in order for $R$ to admit a real positive minimum; in addition, since $\rho=p_{\phi}-p_{z}$ by hypothesis, we must also have $p_{\phi}>0$. If $p_{z}<-\rho$ and/or $p_{\phi}<0$, then condition (\ref{ncc}) is violated and there are no local minima of $R$, with collapse inevitably proceeding to $R=0$, wherein a spindle-like singularity forms. We remark that this case requires at least one of the principal pressures to be non-vanishing, since $p_{z}=p_{\phi}=0\Rightarrow\rho=0$.

\subsection{$\Delta=\rho$}

This case corresponds to isotropic pressure, $\Delta=\rho\Leftrightarrow p_{z}=p_{\phi}$. The acceleration equation at the local extrema reads
\beq
\ddot{R}_{\rm e}=\eta_{-}\left(\eta_{+}\fr{p_{\phi}}{R\rho}+D_{\perp}\psi_{-}\right)-\fr{\eta_{-}\rho}{4R}-R\{\dot{\psi}^{2}_{\Sigma}+(D_{\perp}\psi_{-})^{2}\}. \label{acc2}
\eeq
At the extrema, $\dot{R}_{\rm e}=0$, and, since $\Delta=\rho$, it follows from the junction conditions (\ref{jcz}), (\ref{jcph}) that
\beq
[D_{\perp}\psi]=[\eta]\psi_{,r}^{\Sigma}=\fr{1}{2R}[\eta]\Rightarrow\left.\psi_{,r}^{\Sigma}\right|_{\tau_{\rm e}}=\fr{1}{2R_{\rm e}}.
\eeq
Equation (\ref{acc2}) can then be rewritten as
\beq
\ddot{R}_{\rm e}=\fr{\eta_{-}}{R_{\rm e}}\left(\eta_{+}\fr{p_{\phi}}{\rho}+\fr{\eta_{-}}{4}\right)-\left(R_{\rm e}\dot{\psi}^{2}_{\Sigma}+\fr{\rho\eta_{-}}{4R_{\rm e}}\right).
\eeq
The last summand is manifestly negative, so a necessary condition for $\ddot{R}_{\rm e}>0$ is
\beq
\eta_{+}\fr{p_{\phi}}{\rho}+\fr{1}{4}\eta_{-}>0, \label{cond2}
\eeq
where all quantities are evaluated at $\tau=\tau_{\rm e}$. Clearly, if the WEC holds, the condition above is satisfied. Condition (\ref{cond2}) defines a negative lower bound for $p_{\phi}$,
\beq
p_{\phi}>-\fr{1}{4}\rho\fr{\eta_{-}}{\eta_{+}},
\eeq
which would be incompatible with a violation of the WEC (i.e., with $p_{\phi}<-\rho$) iff $4\eta_{+}\geq\eta_{-}$. By the junction condition (\ref{jcph}), this is equivalent to $\eta^{\rm e}_{+}\geq (R\rho)/3$, where $\eta^{\rm e}_{+}=e^{\psi_{\Sigma}-\gamma_{+}}$. Now, for a given $r_{*}=R(\tau_{*})$, we can rescale the Killing coordinate $z$, such that $\psi^{*}_{\Sigma}=\psi_{\Sigma}(R_{*})=0$; doing this for $R_{\rm e}$, the inequality $\eta^{\rm e}_{+}\geq (R\rho)/3$ becomes
\beq
\gamma_{+}^{\rm e}\leq-\ln(R\rho/3). \label{c3}
\eeq
To examine whether this can at all be verified, we will assume that, in addition to being momentarily static (which, by definition, is the case at the extrema), the configuration is also radiation-free at $\tau=\tau_{\rm e}$, i.e., that $\psi_{,t}=\psi_{,tt}=0$ at $t_{\pm}(\tau_{\rm e})$. With these ansatze, the wave-equation (\ref{weq}) has the trivial solution:
\beq
\psi_{\pm}=\psi_{\Sigma}-k_{\pm}\ln(r/R),
\eeq
where $k_{\pm}$ is a constant. Then, Eqs. (\ref{e1}),(\ref{e2}) yield
\beq
\gamma_{\pm}=\gamma^{\rm e}_{\pm}+k_{\pm}^{2}\ln(r/R).
\eeq
The regularity conditions (\ref{regc}) imply $k_{-}=\gamma_{-}^{\rm e}=0$, thereby fixing the interior solution as $\psi_{-}=\psi_{\Sigma}$ and $\gamma_{-}=0$. From Eqs. (\ref{eta}) and (\ref{jcph}), together with the solution for $\gamma_{-}$, we get
\beq
\gamma^{\rm e}_{+}=-\ln(1-R\rho).
\eeq
Hence, condition (\ref{c3}) would be automatically satisfied (i.e., the WEC would be a necessary condition for collapse to halt) iff
\beq
R\rho\leq\fr{3}{4}.
\eeq
This inequality is stronger than the one imposed by radial nonclosure of space ($R\rho<1$), and as such may or may not hold at $\tau=\tau_{\rm e}$. For this particular case, the approach adopted here cannot show that a violation of the WEC contradicts condition (\ref{cond2}) without additional assumptions for the initial data (this does not, of course, imply that a violation of the WEC necessarily allows for collapse to be halted). This case is nevertheless instructive in that it (i) shows that the WEC is a sufficient condition for collapse to have a chance of being halted, and (ii) provides a self-consistency check: if the WEC holds, then the requirement of radial nonclosure of space is automatically satisfied.

\section{Conclusions}

We studied in detail the formation of trapped surfaces in polarized cylindrical spacetimes with a thin shell of arbitrary matter. We showed that, for arbitrarily large $r$, no trapped surfaces form, either in the vacuum regions, or on the shell, regardless of the matter content of the latter. We studied the limiting behavior of the expansion of a congruence of future-directed outgoing radial null geodesics, and found that they become marginally trapped {\em exactly} at future null infinity. This limiting behavior is due to the interaction between the null geodesics and an orthogonal (locally) plane wave, whose non-vanishing transverse component induces the congruence to start focusing, via a Raychaudhuri-type equation.

The fact that no trapped surfaces form irrespective of the matter content for the shell, {\em could} suggest that this is perhaps an artifact of the simple geometry adopted, in which case it {\em would} not necessarily hold in a fully generic cylindrical spacetime, with non-hypersurface-orthogonal Killing vectors (and whose vacuum regions, because of this, admit two polarizations). However, preliminary calculations~\cite{preli} indicate that this may {\em not} be the case, thus in accord with the spirit of the hoop conjecture.

Even for the simple case of irrotacional dust ($p_{z}=p_{\phi}=0$), an analytical solution for the shell motion in closed form appears impossible to obtain, with the only available fully dynamical solution being numeric~\cite{echeverria93}. Since there is no pressure, total implosion is inevitable, and one can thus derive analytical approximations for the late stages of collapse, when the dynamics is highly relativistic, and the shell asymptotes null dust. Such approximations are valuable for an analytical treatment of the singularity and its structure. In the presence of pressures, however, total implosion is not guaranteed (in fact it was shown {\em not} to occur, in the case of couter-rotating dust~\cite{apostolatos&thorne92}) and, accordingly, one cannot use any approximation for the late stages, since one cannot know {\em a priori} whether and/or when collapse is halted. To be able to make statements about the endstate of collapse without numerical aid one must thus resort to specific classes of models and look for simplifying (but admissable) ansatze.

To examine the effects of matter on the occurrence of singularities, we considered here two classes of matter models, which generalize and complement existing examples in the literature. By using the junction conditions on the shell, together with the assumption of positive energy density, we showed that, for the first class the WEC is a {\em necessary} condition for collapse to halt, and a sufficient condition for collapse to be able to halt in the second matter model. Based on these results, and those of pressureless and counter-rotating dust, we have put forward the mild conjecture that {\em realistic matter is required to prevent singularity formation.} In the conjecture, {\em realistic} is meant to imply that the WEC holds {\em and} at least one of the principal pressures is non-vanishing.

The inclusion of another polarization makes the model mathematically more complicated, but the physics appears to remain largely unchanged (in particular, no singularities seem to form)~\cite{piran&nafier&stark85-86}. Accordingly, we expect the trapped surface results presented here to qualitatively extend to the unpolarized case, as well as the conditions regarding the shell dynamics. Work in this direction is currently underway~\cite{goncalves&moncrief}.

\begin{acknowledgments}
It is a pleasure to thank Vince Moncrief for helpful discussions and comments on the manuscript. This work was supported by FCT (Portugal) Grant SFRH-BPD-5615-2001.
\end{acknowledgments}

\appendix

\section{Non-existence of a global four-dimensional coordinate system}

Let us consider the stress-energy tensor associated with $\Sigma$ expressed as a four-dimensional distribution:
\beq
T^{\mu\nu}_{\Sigma}:=S^{ab}e^{\mu}_{(a)}e^{\nu}_{(b)}|\alpha|\delta(\Phi).
\eeq
Its components are:
\beq
T^{\mu\nu}_{\Sigma}=\rho\eta^{-1}_{\pm}e^{2(\psi_{\Sigma}-\gamma_{\pm})}\delta(\Phi) \mbox{diag}\left(\eta^{2}_{\pm},\dot{R}^{2},p_{z}e^{-2\psi_{\Sigma}},p_{\phi}R^{-2}e^{2\psi_{\Sigma}}\right).
\eeq
The independent non-vanishing components of $G_{\mu\nu}$ are:
\bqa
G_{tt}&=&G_{rr}=-\psi_{,t}^{2}-\psi_{,r}^{2}+r^{-1}\gamma_{,r}, \\
G_{tr}&=&-2\psi_{,t}\psi_{,r}+r^{-1}\gamma_{,t}, \\
%G_{rr}&=&-\psi_{,t}^{2}-\psi_{,r}^{2}+r^{-1}\gamma_{,r}, \\
G_{zz}&=&e^{-2\gamma+4\psi}\left[2(\psi_{,tt}-\psi_{,rr}-r^{-1}\psi_{,r})+\gamma_{,rr}-\gamma_{,tt}-\psi_{,t}^{2}+\psi_{,r}^{2}\right], \\
G_{\phi\phi}&=&e^{-2\gamma}r^{2}\left(\gamma_{,rr}-\gamma_{,tt}-\psi_{,t}^{2}+\psi_{,r}^{2}\right).
\eqa
Einstein's equations give then
\bqa
&&-\psi_{,t}^{2}-\psi_{,r}^{2}+r^{-1}\gamma_{,r}=T_{tt}=T_{rr}, \label{4d1} \\
&&\gamma_{,t}=2r\psi_{,t}\psi_{,r}, \label{4d2} \\
&&\psi_{,tt}-\psi_{,rr}-r^{-1}\psi_{,r}=\fr{1}{2\eta}(p_{z}-p_{\phi})\delta(\Phi). \label{4dweq}
\eqa
In the vacuum regions these equations reduce to the set (\ref{weq})-(\ref{e2}), as expected. On the shell, however, they imply $T^{\Sigma}_{tt}=T^{\Sigma}_{rr}\Rightarrow\eta^{2}=\dot{R}^{2}$, which is only true if [cf. Eq. (\ref{eta})] $\gamma^{\Sigma}_{\pm}\rightarrow+\infty$, but this in turn renders the four-metric singular on $\Sigma$. This simply means that one {\em cannot} use a four-dimensional formulation for the whole spacetime, since the (four-dimensional) Einstein equation $T_{tt}=T_{rr}$ only holds in the vacuum regions. One must therefore adopt two different coordinate systems, for the inner and outer vacuum regions, and match them across $\Sigma$ via the Darmois-Israel junction conditions.

\section{Elementary properties of Bessel functions}

Consider the integral
\beq
I(t,r)\equiv\int^{\infty}_{0} A(\omega)e^{-i\omega t}H_{0}^{(1)}(\omega r)d\omega,
\eeq
where $A(\omega)$ is a complex-valued function, and $H_{n}^{(1)}(x):=J_{n}(x)+iY_{n}(x)$ is a Hankel function of the first kind, and $J_{n}$ and $Y_{n}$ are Bessel functions of the first and second kind, respectively. Partial differentiation of $I$ yields
\bqa
I_{,t}&=&-\int^{\infty}_{0} i\omega A(\omega)e^{-i\omega t}H_{0}^{(1)}(\omega r)d\omega, \\
I_{,r}&=&-\int^{\infty}_{0} \omega A(\omega)e^{-i\omega t}H_{1}^{(1)}(\omega r)d\omega.
\eqa

Now, for $z\gg1$, we have
\beq
H_{n}^{(1)}(z)\sim \sqrt{\fr{2}{\pi z}}e^{i[z-\fr{\pi}{2}(n+\fr{1}{2})]},
\eeq
from which it follows that, for $\omega r\gg1$,
\bqa
I_{,t}&\sim&-i\int^{\infty}_{0} \sqrt{\fr{2\omega}{\pi r}} A(\omega)e^{i[\omega(r-t)-\fr{\pi}{4}]}d\omega, \\
I_{,r}&\sim&-\int^{\infty}_{0} \sqrt{\fr{2\omega}{\pi r}} A(\omega)e^{i[\omega(r-t)-\fr{3\pi}{4}]}d\omega, \nonumber \\
&\sim& i\int^{\infty}_{0} \sqrt{\fr{2\omega}{\pi r}} A(\omega)e^{i[\omega(r-t)-\fr{\pi}{4}]}d\omega.
\eqa
Hence, $I_{,r}\sim -I_{,t}$ for $\omega r\gg1$, thereby justifying Eq. (\ref{asb}).

\section{Newman-Penrose formalism for the analysis of limiting geodesic behavior at \fni}

Here we present a summary of the Newman-Penrose (NP) formalism~\cite{np}, which is particularly useful for a geometrical analysis of the behavior of null congruences. 
We introduce a null tetrad, given by two real null vectors $l^{\mu}$ and $n^{\mu}$, and a complex conjugate pair $m^{\mu}$ and $\bar{m}^{\mu}$, defined such that (i) their only non-vanishing inner products are
\beq
l_{\mu}n^{\mu}=-m_{\mu}\bar{m}^{\mu}=1,
\eeq
and (ii) the metric completness relation holds:
\beq
g_{\mu\nu}=l_{\mu}n_{\nu}+n_{\mu}l_{\nu}-m_{\mu}\bar{m}_{\nu}-\bar{m}_{\mu}m_{\nu}.
\eeq
With these definitions, the Ricci and Weyl tensors can be naturally decomposed in terms of their tetrad components. The Ricci tensor can be decomposed into a scalar component (which gives the curvature scalar) and a Hermitian $(3\times3)$ matrix $\Phi_{AB}$ which represents the trace-free part of the Ricci tensor and satisfies $\Phi_{AB}=\bar{\Phi}_{BA}$. The ten independent components of the Weyl tensor $C_{\alpha\beta\mu\nu}$, which represent the ten degrees of freedom of the gravitational field, are conveniently expressed as five complex {\em NP scalars}. The one relevant for our purposes is
\beq
\Psi_{(0)}=-C_{\alpha\beta\mu\nu}l^{\alpha}m^{\beta}l^{\mu}m^{\nu}.
\eeq
Each NP scalar has a distinct physical interpretation, associated with the presence of gravitational waves along the null directions $l^{\mu}$ and $n^{\mu}$; $\Psi_{(0)}$ denotes a transverse wave component in the $n^{\mu}$ direction.

The behavior of geodesic null congruences (among other things) can be conveniently described in terms of {\em NP spin coefficients}, which are complex linear combinations of the Ricci rotation coefficients associated with the null tetrad. In the present case, the relevant spin coefficients are
\bqa
\theta&=&\nabla_{\nu}l_{\mu}m^{\mu}\bar{m}^{\nu}, \\
\sigma&=&\nabla_{\nu}l_{\mu}m^{\mu}m^{\nu}, \\
k&=&\nabla_{\nu}l_{\mu}m^{\mu}l^{\nu}.
\eqa
They measure the expansion, shear, and deviations from (null) geodesic motion for null rays along $l^{\mu}$, respectively. The evolution of these quantities is given by directional derivatives in the directions of the four tetrad vectors:
\beq
D=l^{\mu}\nabla_{\mu},\; \Delta=n^{\mu}\nabla_{\mu},\; \delta=m^{\mu}\nabla_{\mu},\;\bar{\delta}=\bar{m}^{\mu}\nabla_{\mu},
\eeq
from which one obtains
\bqa
D\theta-\bar{\delta}k&=&\theta^{2}+\sigma\bar{\sigma}-\bar{k}\tau-k(3\alpha+\bar{\beta}-\pi)+\Phi_{00}, \label{ee} \\
D\sigma-\delta k&=&\sigma(\theta+\bar{\theta})+\sigma(\epsilon-\bar{\epsilon})-k(\pi-\bar{\pi}+\bar{\alpha}+3\beta)+\Psi_{(0)}. \label{se}
\eqa
These equations govern the evolution of the expansion and shear along the null $l^{\mu}$ direction.

The natural null tetrad for the Einstein-Rosen metric (\ref{erm}) is (where the metric signature has been changed to $-2$, to conform with the original NP construction):
\bqa
l_{\mu}&=&e^{2(\gamma-\psi)}(\delta^{t}_{\mu}+\delta^{r}_{\mu}), \\
n_{\mu}&=&\fr{1}{2}(\delta_{\mu}^{t}-\delta_{\mu}^{r}), \\
m_{\mu}&=&-\fr{1}{\sqrt{2}}e^{\psi}(i\delta_{\mu}^{z}+re^{-2\psi}\delta_{\mu}^{\phi}).
\eqa
The non-vanishing Ricci components are: 
\beq
\Phi_{00},\;\;\Phi_{02},\;\; \Phi_{11},\;\; \Phi_{22}.
\eeq
The non-vanishing NP scalars are:
\beq
\Psi_{(0)},\;\; \Psi_{(2)},\;\; \Psi_{(4)}.
\eeq
Finally, the non-vanishing NP spin coefficients are (modulo complex conjugation): $\sigma,\lambda,\theta,\mu,$ and $\epsilon$. The relevant ones are:
\bqa
\sigma&=&\bar{\sigma}=\psi_{,t}-\psi_{,r}+\fr{1}{2r}, \\
\theta&=&\bar{\theta}=-\fr{1}{2r}, \\
\epsilon&=&\bar{\epsilon}=\psi_{,r}-\psi_{,t}-\gamma_{,r}+\gamma_{,t}.
\eqa

The relevant evolution equations become then
\bqa
D\theta&=&\theta^{2}+\sigma\bar{\sigma}+\Phi_{00},  \\
D\sigma&=&\sigma(\theta+\bar{\theta})+\Psi_{(0)}.
\eqa

All of the NP objects were computed with the aid of the algebraic package GRTensorII for {\sc Maple} V. Most of the non-zero objects are `too large' to be explicitly written in any convenient way. Since only the fact that they are non-vanishing matters for our purposes (specifically, $\Psi_{(0)}\neq0$), we omit here their explicit form.


\begin{thebibliography}{99}
\bibitem{ftn1}
We note that we refer somewhat loosely to \fni as the locus of events for which the spatial radial coordinate $r$ diverges positively, for a constant value of the retarded Bondi coordinate $u$. A precise definition of \fni in non-asymptotically flat radiative spacetimes is fairly non-trivial, and even different families of static cylinders admit different \fni; see, e.g., Ref.~\cite{ashtekar&bicak&schmidt97} below.
\bibitem{ftn3}
Note that here $n^{\mu}$ is a null tetrad vector, not to be confused with the spacelike unit normal defined in (\ref{sun}).
\bibitem{ftn6}
This condition was explicitly derived in Ref.~\cite{apostolatos&thorne92} for MSRF configurations, and shown to hold throughout collapse.
\bibitem{preli}
A straightforward analysis similar to the one in Sec. III.B, for the case of an unpolarized metric (which contains a non-vanishing cross-term $g_{z\phi}$) reveals that, as in the polarized case, no trapped surfaces form. The analyses of III.A and III.C for the unpolarized case are more cumbersome, and are currently underway (cf. Ref.~\cite{goncalves&moncrief} below).
\end{thebibliography}
\end{document}